\documentclass[conference]{IEEEtran}
\IEEEoverridecommandlockouts
\usepackage{cite}
\usepackage{amsmath,amssymb,amsfonts}
\usepackage{algorithm}
\usepackage{graphicx}
\usepackage{booktabs}
\usepackage{textcomp}
\usepackage{tabularx}
\usepackage{xcolor}
\usepackage{hyperref}
\usepackage{algpseudocode}
\usepackage{cleveref}
\usepackage{authblk}
\usepackage{svg}
\usepackage{xspace}
\usepackage[a4paper, top=19mm, bottom=43mm, left=13mm, right=13mm]{geometry}

\def\ours{\xspace{Web3DB}\xspace}
\def\BibTeX{{\rm B\kern-.05em{\sc i\kern-.025em b}\kern-.08em
    T\kern-.1667em\lower.7ex\hbox{E}\kern-.125emX}}

\title{\ours: Web 3.0 RDBMS for Individual Data Ownership}
\author[$\dagger$]{Shankha Shubhra Mukherjee}
\author[$\dagger$]{Wenyi Tang}
\author[$\dagger$]{Gustavo Prado Fenzi Aniceto}
\author[$\bigstar$]{Jake Chandler}
\author[$\bigstar$]{\\WenZhan Song}
\author[$\dagger$]{Taeho Jung\thanks{*This work was sponsored by NSF under Grant No. OAC-2312973.}}
\affil[$\dagger$]{University of Notre Dame, \textit{Notre Dame, Indiana}}
\affil[$\bigstar$]{University of Georgia, \textit{Athens, Georgia}}

\begin{document}

\maketitle
\begin{abstract}
This paper introduces Web3DB, a decentralized relational database management system (RDBMS) designed to align with the principles of Web 3.0, addressing critical shortcomings of traditional centralized DBMS, such as data privacy, security vulnerabilities, and single points of failure. Several similar systems have been proposed, but they are not compatible with the legacy systems based on RDBMS.
Motivated by the necessity for enhanced data sovereignty and the decentralization of data control, Web3DB leverages blockchain technology for fine-grained access control and utilizes decentralized data storage.  This system leverages a novel, modular architecture that contributes to enhanced flexibility, scalability, and user-centric functionality. Central to Web3DB's innovation is its decentralized query execution, which uses cryptographic sortition and blockchain verification to ensure secure and fair query processing across network nodes. The motivation for integrating relational databases within decentralized DBMS primarily stems from the need to combine the robustness and ease of use of relational database structures with the benefits of decentralization. This paper outlines Web3DB’s architecture, its practical implementation, and the system's ability to support SQL-like operations on relational data, manage multi-tenancy, and facilitate open data sharing, setting new standards for decentralized databases in the Web 3.0 era.
\end{abstract}


\section{Introduction}
In response to the evolving landscape of data management and the imperatives of Web 3.0 principles, we introduce a pioneering decentralized relational DBMS(RDBMS) architecture tailored for the decentralized era. In contrast to conventional centralized architectures prevalent today, our reimagines data management by prioritizing user sovereignty, open data sharing, and decentralized access control. Unlike prior iterations of the web, where data control often gravitated towards centralized entities, Web 3.0\cite{10.1145/3543873.3587583} champions decentralized data networks, offering enhanced security, privacy, and individual data ownership. Our motivation stems from the imperative to mitigate concerns surrounding data privacy, security vulnerabilities, and single points of failure inherent in centralized DBMS architectures.

Let us take the scenario of medical data management. Traditionally, medical records are centralized within hospital systems, posing risks of data breaches and unauthorized access. If a hacker breaches the security of a centralized database, they can potentially access the sensitive medical information of all patients in the system. Data breaches can lead to the exposure of personal health information (PHI), financial details, and other sensitive data, putting patients' privacy at risk. Centralized access can lead to situations where employees with sufficient privileges, either intentionally or accidentally, access patient records without a direct need, breaching patient privacy. Any technical failure, such as server downtime or system malfunction, can deny access to critical patient data when it is most needed, potentially endangering lives.
Decentralized storage allows for patient data to be stored securely across a distributed network, with access keys held by the patient and authorized healthcare providers only, without a single point of failure.
For instance, when a patient undergoes various medical tests, the results, such as pathology reports, must be directly uploaded to the database. The pathologist should have access only to the specific test they conducted, maintaining the privacy of the patient's broader medical records. The patient needs to retain control over their medical data, enabling them to grant or revoke access as needed. Let's say a patient seeks a second opinion; they need to easily share their entire medical history with another doctor without exposing their information to unnecessary risks. This level of control and flexibility is not typically feasible in centralized systems, where data access is managed by the institution rather than the individual.

Central to our approach is recognizing the prevailing gap in decentralized DBMS solutions, particularly in supporting traditional relational databases and SQL-like queries. While the shift towards decentralized databases has gained traction, existing solutions predominantly cater to NoSQL-based DBMS\cite{orbitdb2023}\cite{gundb2023}\cite{couchdb2023}, leaving a void in addressing the complexities of relational data and JOIN operations. Moreover, with traditional relational DBMS continuing to dominate numerous industries\cite{Kammerath2023Relational}, the need for a decentralized architecture compatible with legacy systems becomes imperative.


The research challenges include three primary hindrances.
Our journey through the development of this system is marked by navigating complex challenges. First and foremost, the challenge in the realm of decentralized database management systems (DBMS) lies in the lack of a modular, scalable, and flexible architecture that can adequately cater to varied decentralized data-sharing scenarios and specific user requirements. The evolving landscape of data management, especially in the context of Web 3.0, necessitates a framework that can handle various data ownership models, privacy requirements, and access control mechanisms across disparate and geographically distributed nodes. Each of these elements can vary greatly between different use cases, such as healthcare, finance, or education, adding layers of complexity to the system design. Secondly, enforcing true individual data ownership presents a formidable challenge, given the centralized control typical of traditional DBMS models. In all current data centers, data are possessed and effectively owned by the data centers as they have ultimate decisions on who can access the data. Especially, relational databases are structured with intricate relationships and dependencies between tables and entities. Implementing decentralized access control must account for these complexities, ensuring that permissions and access rights are appropriately managed across the relational model without compromising data integrity. Finally, implementing fine-grained access control in a decentralized environment, essential for multi-tenancy \cite{matthew2014review}, where a single database or table serves multiple tenants or users, proves daunting due to the lack of a central control point and the need for trust and consensus among disparate nodes. Multi-tenancy is relatively straightforward in centralized systems but becomes complex and challenging in decentralized environments.

In this paper, we propose a novel architecture for a Decentralized Relational DBMS, emphasizing a modularized and layered design tailored for Web 3.0. We ensure data ownership by employing a decentralized access control list, managed via smart contracts and blockchain, which empowers users with granular access control and stores the data in a decentralized data storage layer, underscoring our commitment to decentralized data management principles.

Our contributions are multifold:

1) Novel and Modularized Architecture: Our system represents a significant leap forward with its novel, layered, and modularized design, providing a robust framework for decentralized database management. This architecture distributes data across multiple nodes using a decentralized Data Storage Layer, ensuring resilience and eliminating single points of failure. What sets this design apart from state-of-the-art decentralized DBMS is its ability to support relational SQL queries in a decentralized context, bridging the gap between traditional DBMS functionality and the decentralized, distributed nature of Web 3.0 systems.

2) Decentralized Access Control and Data Sovereignty: At the core of our contributions is its implementation of blockchain-based access control, which enforces fine-grained permissions to row-level and ensures user sovereignty over multi-tenant data even in a single table, allowing dynamic and secure access control management. 

3) Decentralized Query Execution: At the core of our decentralized query execution is the cryptographic sortition mechanism. This process employs a secure and unbiased method to select a master node from the network of database engines. The use of cryptographic sortition ensures that the selection is random, fair, and tamper-resistant, facilitated by blockchain technology for added security and verifiability.

4) Open Access and Demonstrable Functionality: The system's transparency is augmented by releasing a fully functional prototype of a decentralized RDBMS based on our architecture: \ours, providing a critical resource for empirical evaluation. This prototype is a demonstrable entity for the academic community and practitioners, permitting direct interaction, testing, and assessment of the system’s efficacy and practical utility. \ours offers comprehensive documentation of its system architecture and functionalities\cite{web3db_runquery2023}\cite{web3db_frontend_github}\cite{web3db_backend_github}.

\ours represents a paradigm shift in database management, offering a cohesive solution to critical challenges in data security, privacy, and decentralization. This paper elucidates our architectural design, system workflow, implementation, experiments, and the potentially transformative impact of \ours, setting a new benchmark for decentralized database systems in the Web 3.0 era.

\section{Related Works}
\noindent \textbf{Distributed DBMS.} The evolution of Distributed Database Management Systems (D-DBMS) has been pivotal in developing scalable and reliable data storage solutions. Early works, such as Özsu et al.\cite{zsu1991DistributedDS}, provide foundational insights into the principles of distributed databases, highlighting their advantages in terms of scalability, reliability, and efficiency over centralized systems. Recent studies, like Corbellini et al. \cite{Corbellini2017PersistingBT}, delve into advancements in distributed systems, focusing on the challenges and solutions in data consistency and fault tolerance.

Several contemporary distributed DBMS architectures have been proposed and implemented. Apache Cassandra \cite{10.1145/1773912.1773922} offers a distributed structure that excels in handling large volumes of data across many commodity servers. Google's Spanner \cite{10.1145/2491245} is another landmark system that combines the benefits of traditional relational databases with the scalability of the NoSQL system. However, optimizing query execution in a decentralized environment is more complex due to the lack of global knowledge about the data distribution and system state. Moreover, traditional distributed databases often have some form of central coordination or master nodes that manage data distribution and query planning.

\noindent \textbf{Decentralized DBMS.} With the advent of Web 3.0, the importance of decentralized databases has gained unprecedented attention. Tapscott et al. \cite{10.5555/3051781}, in their work on blockchain technology, illustrate how decentralized systems underpin the principles of Web 3.0, offering enhanced data security and user sovereignty. These systems, as argued by Swan \cite{10.5555/3006358}, are crucial in the shift towards a more open, interconnected, and user-centric web.


The landscape of decentralized DBMS is still evolving. Technologies like BigchainDB \cite{bigchaindb_whitepaper2023} represent early attempts to integrate blockchain features into database systems, focusing on scalability and decentralized control. However, these solutions predominantly revolve around NoSQL paradigms, as discussed by Senthil et al. \cite{10.14778/3342263.3342632} in their evaluation of blockchain-based databases.

\noindent \textbf{Blockchain-Based Access Control.} Blockchain-based access control mechanisms have gained attention in recent years for their potential to address security and privacy concerns in the Internet of Things (IoT) \cite{10.3389/fdata.2022.1081770}. These mechanisms leverage the decentralized and immutable nature of blockchain to provide integrity and security without relying on a central authority \cite{10.1007/978-981-19-3590-9_16}. Several research studies have explored the use of blockchain for access control in IoT, highlighting its applicability and benefits \cite{Bagga2022}\cite{BashirDar_HamidLone_Naaz_IqbalBaba_Wu_2022}\cite{AgrawalGupta2023}. These studies discuss various access control methods, including certificate-based, certificate-less, and blockchain-based approaches, and compare their pros and cons. 
The concept of decentralized access control in DBMS is a relatively new area of research. The integration of blockchain for access control in databases, as examined by Xu et al. \cite{https://doi.org/10.1002/nem.2089}, offers a glimpse into the potential of immutable, fine-grained access control mechanisms. These studies, however, are in their infancy and have yet to be fully realized in practical, scalable systems.

Despite these advancements, there is a significant research gap in decentralized DBMS, especially concerning systems that support SQL queries and possess robust access control mechanisms. Most existing decentralized databases are based on NoSQL technology \cite{orbitdb2023} \cite{gundb2023}, as highlighted by previous studies, and lack fine-grained access control, which is imperative in the era of Web 3.0. Furthermore, the familiarity and widespread use of SQL in the database community are not addressed in current decentralized systems.

Our work seeks to fill this gap by introducing a decentralized DBMS that not only supports SQL-like queries but also integrates blockchain-based access control, offering a unique combination of user empowerment, data security, and familiarity with SQL. This represents a significant step forward in the field, aligning with the decentralized, user-centric vision of Web 3.0, and it sets a new precedent for future research and development in decentralized database systems.

\section{System Architecture}
\begin{figure}[t]\centering
  \includegraphics[width=0.4\linewidth]{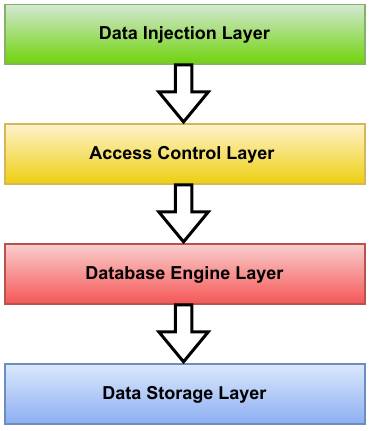}\vspace{-10pt}
  \caption{\ours System Architecture}\label{fig:architecture}\vspace{-10pt}
\end{figure}

The system architecture comprises four layers (\Cref{fig:architecture}): the Data Injection layer, the Access Control layer, the Database Engine layer, and the Data Storage layer. 
Such a layered design provides modularization, and it also allows one to selectively decentralize any layer/component of their choice, thereby providing maximal flexibility such that different trade-offs between decentralization and performance (e.g., throughput and latency) can be sought in different use case scenarios. A modular design is essential to address such diversity in the Web 3.0 domain.

\subsection{Data Injection Layer}
The Data Injection Layer in our proposed architecture serves as the primary interface for user interaction, seamlessly bridging the gap between the end users and the underlying decentralized database management system. This layer 
enables users to issue queries and interact with the system. This layer can be implemented with any form of interface (e.g., web-based interface, command-line interface, and direct APIs). 

Data Injection Layer facilitates API request routing to other layers.
Moreover, the Data Injection Layer plays a crucial role in managing the decentralized authentication mechanisms in our architecture. Users store their unique keys in a decentralized digital wallet, which is pivotal for accessing and interacting with their or others' shared data. These keys are essential for the authorization process, enabling users to allocate or revoke access rights to their data dynamically. The query is encrypted using these keys. Upon query submission, the keys are utilized to communicate with the blockchain-based access control list in the access control layer (\Cref{sec:ac-layer}), retrieve the corresponding data, and ensure secure and authorized data access.

\subsection{Access Control Layer}\label{sec:ac-layer}

The Access Control Layer in our proposed architecture is designed to enforce true individual data ownership and manage access permissions within a decentralized framework. 
At the core of the Access Control Layer is the integration with a permissionless blockchain and an Access Control List(ACL), which serves as an immutable ledger for recording access permissions and ownership details. The Access Control Layer also stores a public log of public keys of all the nodes from the Database Engine layer. This blockchain infrastructure ensures transparency, security, and non-repudiation of access rights, as every transaction and modification in the ACL is recorded and verifiable across the network.

The ACL within this layer maintains a comprehensive record of users' public keys, the associated hashes of the data they own, and a list of hashes for data they are permitted to access. This structured approach to access management allows for:
(I) Verification of Ownership: The ACL cross-references users' public keys with the data hashes to verify ownership, ensuring that only rightful owners or authorized users can access the data.
(II) Granular Access Permissions: The system supports fine-grained access control, where permissions can be specified at the level of individual data items or tables, allowing for nuanced and dynamic access configurations.
(III) Decentralized Validation Process: When a user initiates a query through the Data Injection Layer, their public key and the requested data tables are sent to the Access Control Layer. The blockchain network then validates these requests by cross-checking the ACL entries, ensuring that the user has the necessary permissions.

The permissionless blockchain utilizes a consensus mechanism to authenticate and authorize access requests. This decentralized validation process ensures that:
(I) All network peers agree on the legitimacy of the access request based on the recorded data in the blockchain.
(II) The integrity and consistency of the ACL are maintained across the network, preventing unauthorized access and tampering with access permissions.Once access is verified, the Access Control Layer sends the approval and the query to the Database Engine layer. 

The Access Control layer is also responsible for generating consensus proofs for \textit{cryptographic sortitions} performed in the Database Engine layer (\Cref{sec:data-engine-layer}), which is a protocol run by the database engine nodes to choose a temporary master node for each query to orchestrate it among all decentralized query execution nodes in the Database Engine Layer. The nodes send back their respective calculated \textit{hash}, proof $\pi$ and $j$. The peers in the blockchain verify the sortition using \ref{alg: v_sort}, as used in Algorand \cite{10.1145/3132747.3132757}. After reaching a consensus, the Access Control layer extracts the required hash from ACL and sends it to the master node. After the database engine layer(\Cref{sec:data-engine-layer} executes the entire query, it sends back the updated database hash and query result. Finally, the Access Control layer sends the query result to the Data Injection layer and stores the updated hash on the ACL.

By leveraging blockchain technology, the Access Control Layer in the proposed architecture achieves a decentralized architecture that enhances security, ensures transparency, and provides a robust framework for managing access rights in a distributed environment.
\subsection{Database Engine Layer}\label{sec:data-engine-layer}

The Database Engine Layer is designed to synergize the benefits of decentralized technologies with the robustness and efficiency of traditional DBMS. We incorporate advanced decentralized processing mechanisms, utilizing a combination of cryptographic sortition, gossip network communication, and verifiable random functions.

The Database Engine layer comprises multiple database engine nodes. We maintain a public log of each node's public key along with their weight \textit{$w_i$} (it is described later). We establish a gossip network among the nodes, where each node picks a few random nodes to share messages with. Each node $i$ has its own public and private key ($pk_i, sk_i$), respectively. Every message is signed with the sender's secret key to stop fake messages. Others check this signature to ensure the message's authenticity before forwarding it. To prevent messages from looping in a few nodes only, nodes forward new messages only. Upon receiving a query, the Database Engine Layer broadcasts the request across its gossip network. This network facilitates efficient and rapid dissemination of information, ensuring that all nodes are aware of the incoming query.

In a distributed SQL database management system (DBMS), a master node is the central server that manages data updates and transactions. It coordinates changes to the database and ensures that all the other nodes (often called slave or worker nodes) have the current data. We eliminate this central point in our design by selecting a new temporary master node for each query randomly in a decentralized and fair way.
Cryptographic sortition (\Cref{alg:sortition}), a probabilistic selection process, is employed to choose a master node for the query execution. This process involves a verifiable random function (VRF) executed by each node, producing a random output that determines its likelihood of becoming the master node. The goal is to select the master node in a verifiably random and unbiased way, preventing any node from having disproportionate influence or control over the query processing. 

Sortition is random, so multiple nodes may be selected to be a master node, and the priority determines which master node everyone should adopt. A master node does not keep a private state, except the private key, so it can be easily replaced after each query is executed to mitigate targeted attacks on them. The weight of a user will either be 0 or 1. The user who has served as a worker at least once will have \textit{$w_i$}=1, or else 0. This ensures that new users cannot immediately become master nodes. Also, this technique does not give extra weights to long existing nodes, which ensures that they do not have an advantage and new nodes can become master quickly. The probability that a user will be selected is proportional to \textit{$w_i/W$}. 

\begin{table}[t]
    \centering
    \caption{Frequently used notations}
    \label{notations}\vspace{-10pt}
    \small 
    \begin{tabularx}{\columnwidth}{|X@{\hspace{-120pt}}|X|} 
        \hline
        \textbf{Notation} & \textbf{Definition} \\
        \hline
        $sk_i$ & Secret key of node $i$ \\
        \hline
        $seed$ & Pseudo-random value used for $VRF$ \\
        \hline
        $w_i$ & Weight of node $i$ \\
        \hline
        $W$ & Summation of weight of all the nodes\\
        \hline
        $hash$ & The pseudo-random hash\\
        \hline
        $\pi$ & Proof $\pi$ (from the VRF output) \\
        \hline
        $j$ & It denotes the number of times an individual node has been selected\\
        \hline
    \end{tabularx}\vspace{-10pt}
\end{table}

We have a publicly known random \textit{$seed$}.
A different seed is chosen for each round of query. 
The $seed$ is a pseudo-random value, which does not favor any particular node $i$.
The $seed$ is chosen in the same way as discussed in Algorand \cite{10.1145/3132747.3132757}.
The input query (a string) is denoted by $x$. 
The Verifiable Random Function (VRF) is denoted as \textit{$VRF_{sk}(x)$}, which returns two values: \textit{hash} and proof $\pi$.
The \textit{hash} is indistinguishable from a random string without knowledge of \textit{$sk_i$}.
The pseudo-random hash determines how many sub-nodes are selected. The probability that exactly $k$ out of the $w$ (the node’s weight) sub-nodes are selected follows the binomial distribution, $B(k; w, p) = \binom{w}{k} p^k (1-p)^{w-k}$, where $\sum_{k=0}^{w} B(k;w,p) = 1$.
The proof $\pi$ enables anyone with access to \textit{$pk_i$} to check that the hash corresponds to $x$.
VRF will provide these properties even if \textit{$pk$} \& \textit{$sk$} are chosen by an attacker.
Sortition must select participants according to their weight, ensuring that those with greater contributions have a proportionately higher chance of being chosen.
This is crucial for preventing Sybil attacks.
A nuanced aspect of this process is that individuals with higher weight might be selected multiple times.
To manage this, sortition provides a \textit{$j$} parameter, which denotes the number of times an individual node has been selected.
Being chosen $j$ times means that the user gets to participate as j different “sub-users.”

\begin{algorithm}[t]\label{alg:sortition}
 \caption{\small SORTITION(\textit{sk, seed, w, W, N})}\label{alg: sortition}
 
 \begin{algorithmic}\small
    \State $<$\textit{hash}, $\pi> \leftarrow$ $\textit{VRF}_{sk}$(\textit{seed})
    \State \textit{p} $\leftarrow \frac{w}{W}$;
    \textit{j} $\leftarrow$ 0

    \While {$\frac{hash}{2^{hashlen}} \notin[\sum_{k=0}^{j} B(k;w,p), \sum_{k=0}^{j+1} B(k;w,p)$] }
    \State \textit{j++}
    \EndWhile
    \State \textbf{return} $<$hash,$\pi$, j$>$
 \end{algorithmic}
\end{algorithm}

Both the Algorithm 1 \& 2 are taken from Algorand's\cite{10.1145/3132747.3132757} Algorithm 1 \& 2 respectively. We have only altered how the weights are assigned and removed the parameter $role$. As shown in Algorithm 1, a node performs sortition by computing ⟨\textit{hash},$\pi$⟩ $\leftarrow$ $VRF_sk$ (seed||role).  The pseudo-random hash determines how many sub-users are selected. The number of selected sub-users is public
ly verifiable using the proof $\pi$ (from the VRF output). 
Sortition provides two important properties. First, given a random seed, the VRF outputs a pseudo-random hash value, which is essentially uniformly distributed between 0 and 2\^(hashlen)-1. As a result, users are selected at random based
on their weights. Second, an adversary that does not know \textit{$sk_i$} cannot guess how many times user i is chosen. 

The outputs of the sortition algorithm are sent back to the Access Control layer. The peers in the blockchain of the access control layer wait for a certain amount of time to receive all the outputs and then converge to a consensus using the consensus protocol $BA\star$, as in Algorand\cite{10.1145/3132747.3132757}. The access control layer sends the proof of consensus and the hash of the database table to the selected master node. The master node, upon receiving the data and consensus proof, orchestrates the query processing. It acts as the central coordinator, distributing the workload among other nodes (worker/slave nodes) identified based on the load balancing algorithm. This distributed approach allows for efficient query execution, leveraging the computational resources of multiple nodes in the network. The master node provides the worker nodes with the proof of consensus and the necessary data or table for processing the query. The worker nodes verify the proof of consensus before agreeing to act as a worker node. After each worker node executes its portion of the query, the master node aggregates the results to form the final query output. This process involves collating data from multiple nodes, ensuring the completeness and accuracy of the query response. The updated data or table, along with the query results, are securely encrypted using the master node’s private key with a symmetric-key encryption and then transmitted back to the Access Control Layer. The peers in the blockchain of the Access Control layer verify the signature of the master node and come to a consensus that the selected master node indeed ran the query. The master node also sends a list of all the worker nodes to the Access Control layer. The blockchain updates the weight of all the worker node's weight to 1 on the public log and updates the master node's weight to 0. 

Post-processing, the master node forces all involved worker nodes to clear their caches, eradicating any residual data or records related to the query, thereby maintaining data privacy and security. The master node also clears its cache, ensuring that no sensitive data remains stored on any part of the network after the completion of the query process. We take measures in case the master node behaves in a malicious way by not removing its cache. After the master node finishes query execution, its weight \textit{$w_i$} is reduced to 0. This ensures that this node cannot again become a master node until it serves once as a worker node again. As a result, this node's cache will be forcefully erased when it behaves as a worker node the next time.

\begin{algorithm}[t]\label{alg:verify}
 \caption{\small verifySORTITION(\textit{pk, hash, $\pi$, seed, w, W, N})}\label{alg: v_sort}
 
 \begin{algorithmic}\small
    \If {$\textit{VerifyVRF}_{pk}$(\textit{hash, $\pi$, seed}) is \textit{FALSE}}
    \State \textbf{return} 0
    \EndIf
    \State \textit{p} $\leftarrow \frac{w}{W}$;
    \textit{j} $\leftarrow$ 0

    \While {$\frac{hash}{2^{hashlen}} \notin[\sum_{k=0}^{j} B(k;w,p), \sum_{k=0}^{j+1} B(k;w,p)$] }
    \State \textit{j++}
    \EndWhile
    \State \textbf{return} \textit{j}
 \end{algorithmic}
\end{algorithm}

\begin{figure*}[ht]
\includegraphics[width=\linewidth]{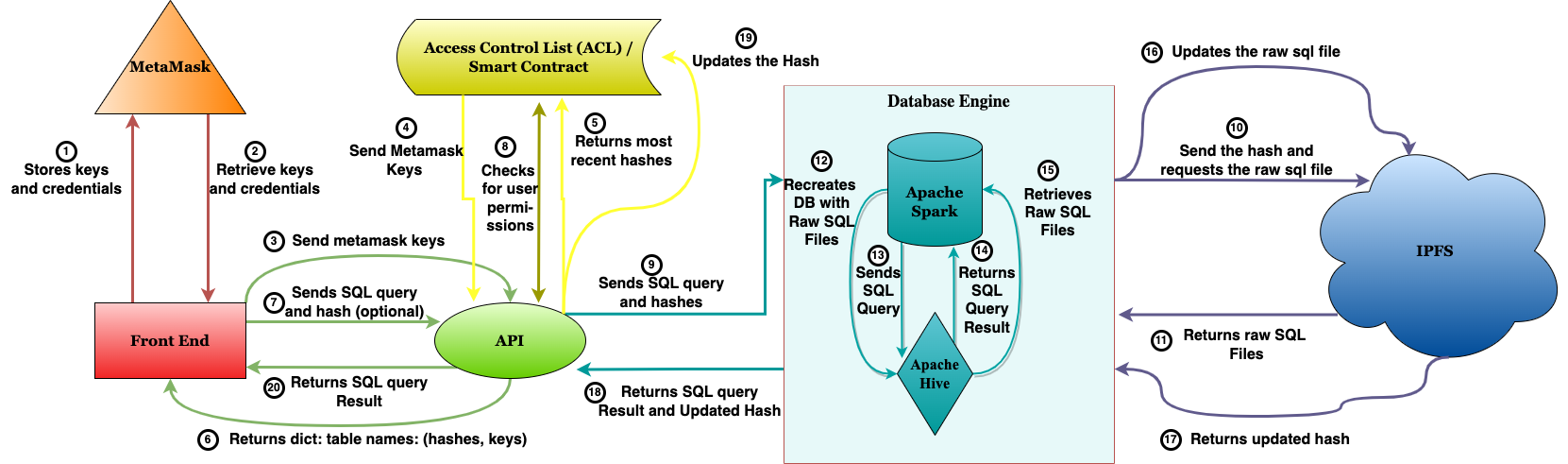}\vspace{-10pt}
  \caption{\ours System Flow}\label{fig:system-flow}
\end{figure*}

This design of the Database Engine Layer ensures fair, transparent, and secure processing of SQL queries in a decentralized database environment.

\subsection{Data Storage Layer}\label{sec:data-storage-layer}
The data storage layer of our system acts as the backbone for the system's decentralized and resilient storage capabilities. We employ a decentralized storage system (e.g. IPFS) for this layer. Instead of using a central server, data are stored across a network of nodes (computers). Each piece of data or file is broken into smaller blocks, which are given a unique identifier (a hash). Nodes store these blocks and share them with others when requested.

The Data Storage Layer is connected to the Access Control Layer, facilitating a smooth interaction where data can be stored and retrieved as required by the database operations. This interaction is vital for the dynamic processing of queries and the real-time updating of data. When a query is received, the peers in the Access Control layer query the Data Storage layer to fetch the required data.  It is retrieved from the Data Storage layer, ensuring that the most current data version is always used. We let the peers instead of the database engine mediate the data retrieval for decentralized access control and data retrieval.

The Data Storage Layer emphasizes data availability and fault tolerance. By leveraging the distributed nature of Decentralized Storage, the system ensures that data is replicated across multiple nodes in the storage network. This redundancy means that even in the event of node failures or network issues, the data remains accessible from other parts of the network, thereby enhancing the system's reliability and uptime.

The Database Engine Layer updates the table back to the Data Storage layer. The updated hash (unique identifier) of the table is sent back in turn to the access control layer, which is then sent back to the Data Injection layer.

\section{Instantiation and Implementation: \ours}

We implement our architecture in a real-world database management system called Web3DB. \ours integrates a sophisticated architectural design with several key components to enable secure and efficient data management in line with Web 3.0 principles. This section delineates the implementation of the system, as depicted in the accompanying \Cref{fig:system-flow}.

\noindent \textbf{Data Injection Layer:} The Data Injection Layer serves as the interactive interface for users, developed using JavaScript to ensure a responsive and engaging user experience. This layer facilitates direct interaction with the Ethereum blockchain via MetaMask, playing a critical role in managing user keys and credentials securely. Upon session initiation by the user, the front-end component retrieves the user’s keys from MetaMask, channeling them to the APIs for further processing. This layer encapsulates the complexities of the underlying infrastructure, offering users seamless access to data and services.

The APIs in the Data Injection Layer is implemented in C\#, which serves as the vital link between the user-facing front end and the backend database engines. It handles incoming API requests, including transmitting user keys to the blockchain’s Access Control List (ACL). Here, smart contracts perform key verification, determine access rights, and retrieve the corresponding data hashes from the data storage layer, ensuring that user interactions are authenticated and authorized in line with the system’s security protocols.

\noindent \textbf{Access Control Layer:}
The Access Control Layer in \ours is instantiated and implemented through a strategic combination of blockchain technology and smart contracts, specifically utilizing Hyperledger Fabric for the underlying infrastructure and logic.

The smart contracts, written in Hyperledger Fabric, define the rules and procedures for verifying user access, managing the ACL, and maintaining the log of public keys of all nodes in the Database Engine layer. These contracts are designed to enforce access policies and ensure that only authorized users can access or modify the data. Hyperledger Fabric’s blockchain network is configured to support the decentralized nature of \ours, facilitating consensus among different nodes regarding access rights and data ownership. When a data access request is initiated through the Data Injection Layer, the system triggers the smart contracts on Hyperledger Fabric. These contracts validate the request against the ACL, utilizing the blockchain ledger to verify user permissions and log the access event.

\noindent \textbf{Database Engine Layer:}
The core of the Database Engine Layer is an integration of Apache Spark and Apache Hive. We use the Boost ASIO library for networking. Apache Spark, often used with Hive for processing, has mechanisms to manage its cache (like unpersisting RDDs or DataFrames) that can be triggered by operations submitted from the master node. To remove data from the cache, Spark provides an unpersist() method, which is called on the Resilient Distributed Dataset(RDD), that was previously cached. 

A distinctive aspect of this layer is its network configuration, employing a gossip protocol to maintain connectivity and synchronization across Hive instances. It is well known that hive is a distributed DBMS. However, integrating cryptographic sortition to select the master node makes it truly decentralized.

Following the query processing, Apache Spark updates the data on IPFS. It communicates the new hash and query results back to the API Layer, ensuring the system's state remains current and reflective of the latest data.

\noindent \textbf{Data Storage Layer:}
The final layer of \ours is built on a public InterPlanetary File System (IPFS) \cite{DBLP:journals/corr/Benet14} network. Opting for a public IPFS network underlines the commitment to decentralization, ensuring distributed, resilient data storage and access across network nodes, fundamental to the architecture of \ours.

\noindent\textbf{Testable Prototype System:}
We have developed a testable prototype of \ours \cite{web3db_runquery2023}\cite{web3db_frontend_github}\cite{web3db_backend_github}, embodying the proposed architecture and demonstrating its feasibility and effectiveness. This prototype serves as a tangible proof of concept for our novel decentralized relational Database Management System (DBMS) suited for the Web 3.0 ecosystem. \Cref{tab:features} is a comparison table of all the features supported by some well-known DBMS.

This is an evolving project that will be regularly maintained and updated with future research. 
For example, we will fully decentralize the API layer using decentralized APIs (APIs). This will bring the system closer to a fully decentralized model, eliminating any central points of failure and further aligning with the principles of Web 3.0. We have provided a link for our prototype system. We have also provided GitHub repository links for our system's front-end and back-end codebases.

\begin{table*}[th]
    \centering
\caption{Comparison to Existing Software/Platform of Distributed/Decentralized DBMS}\vspace{-5pt}
\label{tab:features}
    \begin{tabular}{|c|c|c|c|c|}
         \hline{Distributed/Decentralized}& Full Support of &  Decentralized&  Decentralized& Query \\
         {DBMS}& Relational SQL &  Identity&  Storage& Execution\\\hline
         Google Spanner& $\checkmark$& $\times$& $\times$& Distributed\\
         BigChain DB& $\times$& $\checkmark$& $\checkmark$& Distributed\\
         IceFire DB& $\times$& $\times$& $\checkmark$& Distributed\\
         OrbitDB& $\times$& $\checkmark$& $\checkmark$& Distributed\\
         GunDB& $\times$& $\times$& $\checkmark$& Distributed\\
         \ours (Ours)& $\checkmark$& $\checkmark$& $\checkmark$& Decentralized \\\hline
         
    \end{tabular}\vspace{-10pt}
\end{table*}

\section{Experiments \& Results}

This section aims to evaluate the performance of Web3DB. The TPC-H benchmark\cite{10.1007/978-3-642-30507-8_20}, a standard for evaluating the performance of DBMS, was chosen for testing. In decentralized systems, execution time is crucial for evaluating scalability. By observing how execution times vary with changes in the number of nodes, data volume, or query complexity, developers can understand the scalability characteristics of the system. Moreover, Execution time is a fundamental metric for benchmarking the performance of a DBMS.

\subsection{Experiment Specifications}
We imported TPC-H data with a scale factor of 10 (approximately 10 GB), ensuring a standardized dataset for comparison. The tests were conducted across 1, 2, 5, 10, and 18 parallel nodes to evaluate the system's scalability and query processing efficiency. We did not scale up beyond 18 nodes due to budget and configuration complexity issues. Each node represented an instance of the database engine within the network.

\begin{table}[t]
    \centering
\caption{TPC-H Benchmark test for execution time(Seconds)}\vspace{-5pt}
\label{tab:my_label}
    \setlength{\tabcolsep}{2pt} 
\renewcommand{\arraystretch}{1} 
    \begin{tabular}{|c|c|c|c|c|c|}
         \hline TPC-H Query&  1 Node&  2 Nodes&  5 Nodes&  10 Nodes& 18 nodes\\\hline
         Query 1&  38.02&  35.86&  25.61&  16.32& 13.71\\
         Query 2&  32.16&  30.21&  21.81&  14.78& 12.10\\
         Query 3&  36.87&  34.68&  25.02&  16.41& 13.17\\
         Query 4&  34.64&  32.49&  22.96&  14.64& 12.44\\
         Query 5&  52.85&  49.65&  35.48&  19.10& 15.42\\
         Query 6&  35.57&  33.62&  22.85&  15.22& 13.30\\
         Query 7&  37.10&  34.88&  24.72&  16.39& 13.55\\
         Query 8&  36.89&  34.57&  24.77&  16.56& 13.53\\
         Query 9&  35.81&  33.65&  23.62&  16.01& 13.96\\
         Query 10&  35.83&  33.38&  22.43&  14.13& 12.69\\
         Query 11&  36.53&  34.83&  22.99&  14.22& 13.26\\
         Query 12&  37.70&  35.15&  25.05&  16.41& 13.13\\
         Query 13&  35.71&  33.48&  23.10&  15.45& 12.45\\
         Query 14&  36.90&  34.47&  22.87&  14.33& 13.18\\
         Query 15&  42.65&  40.12&  27.40&  18.80& 14.98\\
         Query 16&  30.81&  28.65&  19.66&  11.62& 11.14\\
         Query 17&  53.14&  50.25&  33.78&  22.81& 16.35\\
         Query 18&  49.08&  45.70&  32.40&  24.22& 15.87\\
         Query 19&  36.70&  34.51&  25.13&  16.97& 13.00\\
         Query 20&  37.05&  34.85&  24.55&  15.95& 13.51\\
         Query 21&  43.64&  40.82&  28.04&  19.22& 15.22\\
         Query 22&  33.82&  31.81&  21.59&  14.74& 12.24\\\hline
    \end{tabular}
\end{table}

AWS EC2 instances were strategically deployed across multiple Availability Zones within several AWS regions to simulate a distributed environment and ensure high availability and fault tolerance. The regions include us-east-1, us-east-2, us-west-2, eu-west-2, eu-west-3, eu-central-1, ap-south-1, ap-northeast-2, ap-southeast-1. Each of these database engine nodes also serves as a public IPFS node by running an IPFS daemon instance (connected to the public IPFS network). This setup also helps in evaluating the system’s performance across a geographically distributed network, reflecting a realistic scenario where nodes are not centrally located. Each instance was configured with a suitable Linux distribution, necessary security settings, and network configurations to enable seamless inter-node communication and secure data transfers. 

The TPC-H queries, consisting of 22 standardized queries, were executed on each node configuration. This size is significant enough to test the scalability and performance of the database system under non-trivial loads. The TPC-H benchmark uses a standardized schema that includes tables like Customer, Orders, LineItem, etc., with predefined relationships. The data distribution within these tables is deliberately skewed for certain columns to mimic real-world scenarios and test the DBMS's capability to handle uneven data distributions effectively.
These queries are designed to cover a wide range of DBMS functionalities, including join operations, aggregations, and nested queries. 
The TPC-H benchmark consists of 22 queries, which vary significantly in complexity and type. The complexity of these queries helps in evaluating the performance of the database system across different types of data operations.
The execution times for each query were recorded in all node setups. 
Throughout the testing phase, care was taken to maintain environmental consistency to ensure that the results were attributable solely to the system’s performance and not external factors.

\subsection{Experiment Results}

The experimental results are shown in {\Cref{tab:my_label}}.
The results indicated a clear trend: as the number of nodes increased, the query execution times decreased. 
This improvement can be attributed to the distributed nature of the system, where parallel processing and efficient workload distribution played a significant role.
The results demonstrate approximately a 6\% improvement in execution times with two nodes, 20\% with three nodes, up to 34\% with five nodes, 58\% with ten nodes, and 64\% with 18 nodes. We tried to scale the system up to 20 nodes, but the network was crashing every time on AWS. This shows that it is challenging to scale a decentralized network, and further optimization of the code is necessary.

Due to the constraints imposed by the page limit for this paper, we were only able to present the TPC-H benchmark test experiment as the sole demonstration of Web3DB's performance and capabilities. This focused approach allowed us to provide a detailed and comprehensive analysis within the available space. The experiment with the TPC-H benchmark on the Web3DB system successfully demonstrated the system's scalability and efficiency in processing complex SQL queries in a decentralized environment. The results validate the effectiveness of the system's architecture and its potential in managing large-scale data with the principles of Web 3.0. 

\section{Discussion \& Future Work}
A key assumption in our design is the stability of the database engines, which are presumed not to go offline unexpectedly. This stability is crucial for the reliability, throughput, and consistency of the system. 
Additionally, according to the CAP Theorem, which states that a distributed system can only fully satisfy two out of three properties—Consistency, Availability, and Partition Tolerance—our system prioritizes Consistency and Availability (CA).
This means that while we prioritize consistency and availability, the system may be less resilient to network partitions compared to CP or AP systems.
The overhead of data transfer between the database engine and the IPFS network increases execution time significantly. This is due to the inability to execute queries directly on IPFS, which uses the Content Addressable aRchive (CAR) format, which is an open problem.

Looking forward, there are several avenues for enhancing our proposed architecture. To bolster security, we plan to integrate more sophisticated encryption techniques. This would enhance data privacy and security, making the system robust against potential cyber threats. Given our system's current alignment with the CA paradigm, exploring ways to improve partition tolerance without significantly compromising consistency or availability would be a valuable area of research. This could involve developing algorithms or mechanisms that enable the system to maintain operations during network partitions.

\section{Conclusion}
In this paper, we have presented a novel decentralized RDBMS architecture designed for the evolving Web 3.0 landscape. 
Our system addresses several critical challenges in the realm of database management, particularly in the context of decentralization and the need for fine-grained access control. 
Our system's layered and modular architecture significantly contributes to the field. It demonstrates how traditional SQL-based query processing can seamlessly integrate with decentralized storage solutions like IPFS. 

The system's ability to handle SQL-like queries on relational data, support multi-tenancy, run decentralized query execution, and facilitate open data sharing sets new benchmarks for functionality and performance in decentralized database systems. The practical implementation of Web3DB, demonstrated through a fully functional prototype, showcases the viability and efficiency of this novel system in real-world applications.

As Web3DB continues to evolve, future research will focus on further decentralizing the system's components, enhancing scalability, and improving the consensus mechanisms across the network. 

\bibliographystyle{ieeetr}
\bibliography{citation.bib}

\end{document}